\documentclass[draft]{epl2}
\usepackage{color,soul} 
\usepackage{mathtools}
\usepackage{float}
\usepackage{graphicx}
\usepackage{amsmath}
\usepackage{amsthm}
\usepackage{amssymb}
\usepackage{multirow} 
\usepackage[normalem]{ulem}
\usepackage{lineno}

\title{Application of Haldane's statistical correlation theory in classical systems}
\author{Projesh Kumar Roy\inst{1}}
\shortauthor{P. K. Roy \etal}
\institute{                    
  \inst{1} Department of Chemistry, National Institute of Technology Tiruchirappalli, Tanjore Main Road, NH83, Tamil Nadu 620015, India.
  \email{projesh@nitt.edu}
}

\date{today}
\abstract{This letter investigates the application of Haldane's statistical correlation theory in classical systems. A modified statistical correlation theory has been proposed by including non-linearity in the form of an exponent into the original theory of Haldane. The dependence of the statistical correlation on indistinguishability is highlighted. Using this modified theory, a quasi-classical derivation of intermediate statistics is shown where indistinguishability can be introduced into distinguishable systems in the form of a statistical correlation. The final result is equivalent to the classical fractional exclusion statistics (CFES), which was derived earlier using a purely classical route. An extended non-linear correlation model based on power series expansion is also proposed, which can produce various intermediate statistical models.}

\begin{document}

\maketitle

%%%%%%%%%%%%%%%%%%%%%%%%%%%%%%%%%%%%%%%%%%%%%%%%%%%%%%%%%%%%%%%%%%%%%%%%%%%%%%%%
\section{Introduction}

In the past decades, there were several attempts~\cite{Gentile_NuovoCimento_1940, Tsallis_JStatPhys_1988, March_Sung_PhysChemLiq_1993, March_PhysChemLiq_1997, Acharya_Swamy_JPhysA_1994} to unify all fundamental statistical theories, i.e., Fermi-Dirac (FD) statistics, Bose-Einstein (BE) statistics, and Maxwell-Boltzmann (MB) statistics. A comprehensive approach for such a unified theory was taken by Haldane~\cite{Haldane_PhysRevLett_1991}. In a nutshell, the postulate of Haldane states that the term {\it statistical correlation} can be described as a reduction of the total available degenerate states ($\tilde{g_i}$) at $i^{\text{th}}$ energy level ($\epsilon_i \in [0,\infty]$) due to the change in the population ($\Delta n_j$)  of the building blocks of the system at different energy levels, $\epsilon_j$, including self interaction ($i=j$). The nature of the reduction was assumed to be linear, which resulted in the following equation~\cite{Wu_PhysRevLett_1994},

\begin{equation}
\tilde{g_i} = g_i - \sum_j \gamma_{ij} \Delta n_{j}
\label{eqn:haldane}
\end{equation}

In the subsequent letter, the word \textit{particle} will be used to denote the building blocks of a system. Eq.~\ref{eqn:haldane} is written for a single species of particles, but can be extended for different species as well~\cite{Haldane_PhysRevLett_1991}. If the particle numbers at all energy levels are allowed to change from $0$ to $n_j$, then $g_i$ is the number of ideal degenerate states at $\Delta n_j = 0$. $\gamma_{ij}$ is a parameter which controls the correlation effects originating from different energy levels. Historically, this theory originated from the context of fractional statistics, which is generally observed in quantum systems of indistinguishable particles. Applying this idea to Bosonic systems, Wu~\cite{Wu_PhysRevLett_1994} was able to derive an equation for \textit{intermediate} statistics which interpolates between Bose-Einstein and Fermi-Dirac statistics. However, Haldanes's theory itself doesn't explicitly need the particles to be indistinguishable objects. The primary condition for this theory needs to be valid is that $g_i$ is independent of the properties of the $n_j$ particles, which is true for both distinguishable or indistinguishable particles.

Application of Haldane-Wu (HW) statistics can be found in the field of 2D-anyones~\cite{Veigy_Ovury_ModPhysLettA_1995_1, Veigy_Ovury_ModPhysLettA_1995_2}, Laughlin liquids~\cite{Camino_Goldman_PhysRevB_2005, Arovas_Wilczek_PhysRevLett_1984}, adsorption of polyatomic molecules~\cite{Riccardo_Roma_PhysRevLett_2004, Riccardo_Pastor_AppSurfSci_2005, Cerofolini_JPhysA_2006, Davila_Pastor_SurfSci_2009, Davila_Pastor_JChemPhys_2009, Fernandez_Pastor_Langmuir_2011, Fernandez_Pastor_ChemPhysLett_2014, Riccardo_Pasinetti_PhysRevLett_2019}, etc. A similar statistical theory was proposed by A. Polychronakos--namely the Polychronakos statistics (AP)~\cite{Polychronakos_PhysLettB_1996, Polychronakos_NuclPhysB_1996, Chung_Hassanabadi_ModPhysLettB_2018, Hoyuelos_PhysicaA_2018}-- which avoids the negative probability problem in HW statistics~\cite{Nayak_Wilczek_PhysRevLett_1994, Polychronakos_PhysLettB_1996, Polychronakos_NuclPhysB_1996, Chaturvedi_Srinivasan_PhysicaA_1997, Chaturvedi_Srinivasan_PhysRevLett_1997, Murthy_Shankar_PhysRevB_1999}. Various other forms of intermediate statistics also exist in the literature~\cite{Niven_EurPhysJB_2009, Niven_Grendar_PhysLettA_2009, Ourabah_Tribeche_PhysRevE_2018, Abutaleb_IntJTheoPhys_2014, Yan_PhysRevE_2021}. Such models can be termed quantum fractional exclusion statistics (QFES) models as they inherently assume the quantum nature of the systems. 

Recently, a theory of intermediate statistics based on classical MB statistics--namely, the classical fractional exclusion statistics (CFES)~\cite{Roy_PhysRevE_2022}--was derived using the maximum entropy (MaxEnt) methods, where three major constraints were used along with the corresponding Lagrange parameters during the derivation: (i) constant energy, (ii) constant particle number, and (iii) a special constrain, which can be written as,

\begin{equation}
S_a = \sum_i \Bigg (\frac{n_i^{a}}{g_i^{a-1}} \Bigg)
\label{eqn:constraint}
\end{equation}

The resulting energy distribution for dilute systems ($g_i >> n_i$) can be written as,

\begin{equation}
n_i = g_ie^{-(\alpha+\beta\epsilon_i)}\Bigg\{1 - ab\Bigg(\frac{n_i}{g_i}\Bigg)^{(a-1)}\Bigg\}
\label{eqn:orig_CFES}
\end{equation}

Where $a$ is a positive non-zero integer and $b$ is essentially a Lagrange parameter related to Eq.~\ref{eqn:constraint}. Interestingly, it was possible to recover FD and BE statistics at $a=2$ from Eq.~\ref{eqn:orig_CFES} using these constraints for dilute systems ($g_i >> n_i$). Thus, due to the special constraint in Eq.~\ref{eqn:constraint}, the overall energy distribution showed a quantum-like behaviour, without any prior assumption of the same. 

However, the origin of this special constraint is still not clear, which was proposed to be related to the indistinguishablility of the particles. Previously in reference~\citenum{Roy_PhysRevE_2022}, I theorized that the factor $S_a$ can be related to the indistinguishability of the particles as it acts as a mathematical bridge between the MB and BE statistics as, 

\begin{equation}
S_{\text{BE}} = S_{\text{MB}} + \Bigg [ \sum_{a=0}^{\infty} f_aS_a - (N\ln N + G)\Bigg ]
\end{equation}

where $S_{\text{BE}}$ and $S_{\text{MB}}$ are the entropy of the BE and MB statistics, respectively. $N$ and $G$ are the overall number of particles and number of degenerate states, respectively. $f_a$'s are the coefficients of logarithmic expansions. Note that, MB and BE statistics differ only in the aspect of indistinguishability.

In this letter, I further explore the origin of the special factor $S_a$, and how it is related to the concept indistinguishability. I show that a non-linear form of Haldane\rq{s} theory can be used to highlight the relationship between the statistical correlation and indistinguishability, which by extension, clarifies the relationship between Eq.~\ref{eqn:constraint} and indistinguishability. Using this modified Haldane's theory, an alternate quasi-classical derivation was used to arrive at CFES-equivalent intermediate statistics. 

The letter is organized as follows: In first section, non-linear modification are applied to Eq.~\ref{eqn:haldane}. In second section, I describe a self-correlating quasi-classical system following the modified Haldane's theory. In third section, effects of extended non-linear modifications using power series expansion is shown. Lastly, I conclude with an outlook. 

%%%%%%%%%%%%%%%%%%%%%%%%%%%%%%%%%%%%%%%%%%%%%%%%%%%%%%%%%%%%%%%%%%%%%%%%%%%%%%%%
\section{Non-Linear Statistical Correlation Theory}
\label{sect:zero_moment_theorem}

Haldane's statistical correlation theory is based on two inherent assumptions: Firstly, the total correlation effect at $\epsilon_i$ level originating from each particle at $\epsilon_j$ level is same, i.e. $\gamma_{ij}$ is a function of $\epsilon$'s only and not particle properties. Secondly, the correlation effect doesn't depend on the individual population of the $n_j$ particles at $k^{\text{th}}$ degenerate state of $\epsilon_j$ level ($n_{jk}$) since the overall probability distribution of the particles at different degenerate states ($p(n_{jk})$) are {\it equal-a-priory}. As shown in Eq.~\ref{eqn:haldane}, Haldane's theory proposes that the total statistical correlation effect is linear with respect to $n_j$. In this section, I wanted to investigate, how Eq.~\ref{eqn:haldane} will change if the total statistical correlation is non-linear.

I achieve this task by removing and then re-imposing the {\it equal-a-priory} criteria to a set of particles of single species. I assume that particles are unevenly distributed in different degenerate levels, and therefore, the net contribution to the statistical correlation from $\epsilon_j$ level to $\epsilon_i$ level may depend on $p(n_{jk})$: thus removing the {\it equal-a-priory} probability condition. The overall contribution of the statistical correlation to level $\epsilon_i$ is weighted by correlation factors $\gamma_{ij}$, which is same for all degenerate states. Then, I impose non-linearity to the net correlation contribution originating from a certain $g_{jk}$ state, by raising an exponent $m$ to the population $n_{jk}$ at that state. Thus, the resulting equation takes the shape,

\begin{equation}
\tilde{g_i} = g_i - \sum_j\sum_k \gamma_{ij} n_{jk}^m
\label{eqn:nonlinear_haldane}
\end{equation}

where I assume particle numbers at all energy levels are allowed to change from $0$ to $ n_j$, i.e., $\Delta n_j = n_j$. Next, I wrap-up this model by reimposing {\it equal-a-priory} probability condition and state that, the probability of finding a particle at any of the $g_{jk}$ degenerate states of an energy level $\epsilon_j$ is equal; i.e., $p(n_{jk}) \equiv n_j/g_j $. For such a system, one can state that, the variance ($\sigma_2$) of the $p(n_{jk})$ function, which is uniformly distributed over $g_{jk}$ degenerate states, is zero. Subsequently--as well known in probability theory~\cite{Papoulis_2002}--all higher order central moments will also be zero, which can be written as,

\begin{equation}
\sigma_m = \frac{\sum_k n_{jk}^m}{g_{j}} - \left(\frac{\sum_k n_{jk}}{g_{j}}\right)^m = 0 
\label{eqn:std_dev}
\end{equation}

which leads to,

\begin{equation}
\sum_k n_{jk}^m = \frac{n_j^m}{g_j^{m-1}}
\label{eqn:std_dev_summand}
\end{equation}

Clearly, $m$ has to be an integer to satisfy the condition in Eq.~\ref{eqn:std_dev_summand}. I will show later that $m$ and $\gamma$ factors become dependent on each other for intermediate statistics. We can now fully eliminate the $k$-summand in Eq.~\ref{eqn:nonlinear_haldane} by substituting Eq.~\ref{eqn:std_dev_summand} to Eq.~\ref{eqn:nonlinear_haldane} as,

\begin{equation}
\tilde{g_i} = g_i - \sum_j \gamma_{ij} \bigg(\frac{n_j^m}{g_j^{m-1}}\bigg)
\label{eqn:nonlinear_haldane_modified}
\end{equation}

One can see that $S_a$-equivalent factor has appeared on the right hand side of Eq.~\ref{eqn:nonlinear_haldane_modified} due to the non-linear treatment of Haldane's theory. It is straightforward to show that Eq.~\ref{eqn:haldane} can be recovered from Eq.~\ref{eqn:nonlinear_haldane_modified} at the linear correlation case at $m=1$. Note that, the linear correlation case exerts a maximum correlation effect on $g_i$. In the following section, I show that the application of Eq.~\ref{eqn:nonlinear_haldane_modified} to classical MB statistics directly gives rise to the CFES-equivalent equation. 

%%%%%%%%%%%%%%%%%%%%%%%%%%%%%%%%%%%%%%%%%%%%%%%%%%%%%%%%%%%%%%%%%%%%%%%%%%%%%%%%
\section{Self-Correlating Classical System}
\label{sect:self_corr}

The non-linear correlation theory can certainly be used to modify the microstate counting formula of BE statistics ($W_{\text{BE}}$) and derive different forms of QFES, as described in reference~\citenum{Wu_PhysRevLett_1994}. As I have stated earlier, Haldane's theory doesn't explicitly need the particles to be indistinguishable; it might as well be applied to distinguishable systems. Previously, we have investigated the effect of statistical correlation in classical 2D-silica model~\cite{Roy_Heuer_PhysRevLett_2019}, where neighbouring silicate rings are correlated to each other via angular strain~\cite{Roy_Heuer_JPhysCondMat_2019}. Using a simple two-state classical model, we have shown that~\cite{Roy_Heuer_PhysRevLett_2019} statistical correlation can successfully describe the silicate ring distributions in 2D-silica. Following a similar line of thought, I apply the modified statistical correlation theory in Eq.~\ref{eqn:nonlinear_haldane_modified} to the distinguishable particles and investigate the changes in the microstate counting in classical MB statistics ($W_{\text{MB}}$) in this section. 

Let's assume that a single species of distinguishable particles is present in the system which are themselves involved in statistical correlation obeying Haldane's theory. Let's denote their un-optimized population with a slightly different notation, $\nu_i$. Next, I reduce the actual number of degenerate states all all energy levels in this system following Eq.~\ref{eqn:nonlinear_haldane_modified}. Note that, this derivation is different from reference~\citenum{Roy_PhysRevE_2022}, as the particles are assumed to have statistical correlation {\it a priory}. Using Eq.~\ref{eqn:nonlinear_haldane_modified}, I can show that $W_{\text{MB}}$ takes the form,

\begin{equation}
W_{\text{MB}} = W_{\text{MB}}^{\text{id}}\prod_i \Bigg\{1-\frac{1}{g_i}\sum_j\gamma_{ij}\Bigg(\frac{\nu_j^m}{g_j^{m-1}}\Bigg)\Bigg\}^{\nu_i}
\label{eqn:microstates_modified}
\end{equation}

where $W_{\text{MB}}^{\text{id}} = \prod_i g_i^{\nu_i}/\nu_i!$ is the ideal number of the microstates for distinguishable classical systems. At a dilute condition; i.e., $\nu_i/g_i \sim 0$; one can derive the corresponding optimized particle distribution in the following manner. Using MaxEnt method, one can maximize $W_{\text{MB}}$ with respect to $\nu_i$ to get the optimized population, $\nu_{i, \text{max}} \equiv n_i$, as,

\begin{equation}
\begin{aligned}
\frac{\partial \ln W_{\text{MB}}}{\partial \nu_i} &= \frac{\partial \ln W_{\text{MB}}^{\text{id}}}{\partial \nu_i} 
																	    +  \ln \Bigg\{1-\frac{1}{g_i}\sum_j\gamma_{ij}\Bigg(\frac{\nu_j^m}{g_j^{m-1}}\Bigg)\Bigg\} \\
																	    & -  \frac{m\gamma_{ii}\nu_i^{m}/g_i^{m-1}}{\{g_i-\sum_j\gamma_{ij}(\nu_j^m/g_j^{m-1})\}} \\
																	   &= 0
\end{aligned}
\label{eqn:maxent}
\end{equation}

which leads to,

\begin{equation}
\begin{aligned}
n_i &= g_ie^{-(\alpha+\beta\epsilon_i)}\Bigg\{1-\frac{1}{g_i}\sum_j\gamma_{ij}\Bigg(\frac{n_j^m}{g_j^{m-1}}\Bigg)\Bigg\} \\
	   &\exp\Bigg\{-\frac{m\gamma_{ii}n_i^{m}/g_i^{m-1}}{g_i-\sum_j\gamma_{ij}(n_j^m/g_j^{m-1})}\Bigg\} \\
	   &\approx \tilde{g_i}e^{-(\alpha+\beta\epsilon_i)}\Bigg\{1-m\gamma_{ii}\Bigg(\frac{n_i^m}{\tilde{g_i}g_i^{m-1}}\Bigg)\Bigg\} \text{ ; using $g_i >> n_i$}\\
	   &= g_ie^{-(\alpha+\beta\epsilon_i)}\Bigg\{\frac{\tilde{g_i}}{g_i}-m\gamma_{ii}\Bigg(\frac{n_i^m}{g_i^m}\Bigg)\Bigg\}
\end{aligned}
\label{eqn:CFES_modified}
\end{equation}

where $\alpha=\beta\mu$ and $\beta=(1/k_{\text{B}}T)$ are the Lagrange parameters related to the chemical potential ($\mu$) and temperature ($T$) of the system. If we assume that it is a self-correlating system, we only need to take into account the diagonal elements, i.e., $\gamma_{ij} \equiv \gamma\delta_{ij}$ ($\delta_{ij} = 0$ for $i \neq j$ and $\delta_{ij} = 1$ otherwise). Then, Eq.~\ref{eqn:CFES_modified} reduces to a CFES-equivalent equation as,

\begin{equation}
n_i = g_ie^{-(\alpha+\beta\epsilon_i)}\Bigg\{1-(m+1)\gamma\Bigg(\frac{n_i^m}{g_i^m}\Bigg)\Bigg\}
\label{eqn:CFES}
\end{equation}

where $\gamma$ corresponds to the third Lagrange parameter and $m$ is the exponent~\footnote{Note that, the exponent $m$ is shifted by 1 in this derivation method as compared to Eq.~\ref{eqn:orig_CFES}} described in Eq.~\ref{eqn:orig_CFES}~\cite{Roy_PhysRevE_2022}. Consequently, one can derive the $S_a$-equivalent special constraint for the population distribution used in Eq.~\ref{eqn:orig_CFES} in a retrospective way as,

\begin{equation}
S_m = \sum_i \Bigg (\frac{n_i^{m+1}}{g_i^m} \Bigg)
\label{eqn:constraint_new}
\end{equation}

For $\gamma > 0$ case (abbr: $\gamma^+$), the relationship between $\gamma$ and $m$ satisfy the equation,

\begin{equation}
n_{\text{c}} = \Bigg\{\frac{1}{(m+1)\gamma^+}\Bigg\}^{1/m}
\label{eqn:max_occupancy}
\end{equation}

where $n_{\text{c}}$ is the maximum occupancy numbers at all available degenerate states. An interesting question comes up about the limits of $m$. In the previous section, I have already shown that $m$ has to be an integer to satisfy Eq.~\ref{eqn:std_dev_summand}. If one only considers Eq.~\ref{eqn:CFES}, $m=[-1,0,\infty]$ will produce traditional MB statistics, as proved in reference~\cite{Roy_PhysRevE_2022}. However, if one considers Eq.~\ref{eqn:nonlinear_haldane_modified}, $m=0$ system become unphysical at $\gamma \to \infty$ because of Eq.~\ref{eqn:max_occupancy}, which will give $\tilde{g}_i \to -\infty$. Similarly, for $m=-1$, $\tilde{g}_i \to \infty$ at $n_i \to 0$. Therefore, if one considers the quasi-classical derivation of CFES under the modified Haldane's theory, the limit of $m$ is $\infty \geq m \geq 1$. Here, $m \to \infty$ and $\gamma \to 0$ system is the only allowed system where MB statistics can be recovered. Thus, Eq.~\ref{eqn:CFES} is an unifying statistical theory that encompasses properties MB ($m = \infty, \gamma = 0$), FD ($m = 1, \gamma = 1$), and BE ($m = 1, \gamma = -1$) statistics, as well as their intermediate statistics.

Thus, my analysis show that applying statistical correlation among distinguishable particles, I can recover all forms of quantum statistics which are generally derived using FD and BE microstate counting methods for indistinguishable particles. Therefore, this work highlights the interdependence of statistical correlation and indistinguishability, and by extension, it also describes how the special constraint $S_a$ in Eq.~\ref{eqn:constraint} is linked to indistinguishability of the particles, supporting my earlier theory. Essentially, I show an alternate path for introducing indistinguishability in a classical distinguishable system, other than the various microstate counting method.

%%%%%%%%%%%%%%%%%%%%%%%%%%%%%%%%%%%%%%%%%%%%%%%%%%%%%%%%%%%%%%%%%%%%%%%%%%%%%%%%
\section{Extended Non-Linear Models}
\label{sect:extended}

It might be possible to use functions other than the simple exponent in Eq.~\ref{eqn:nonlinear_haldane_modified} to introduce non-linearity in the statistical correlation. However, the elimination of the $k$-summand in Eq.~\ref{eqn:nonlinear_haldane} is necessary to maintain realistic energy distributions in any non-linear model. The exponent model in Eq.~\ref{eqn:nonlinear_haldane} was perhaps the simplest where such elimination was easy because of Eq.~\ref{eqn:std_dev_summand}. A more general method to introduce non-linearity--which will also ensure the elimination of the $k$-summand--is to use a power series expansion of $n_{jk}$ as,

\begin{equation}
\tilde{g}_j = g_j - \sum_j  \sum_k \gamma_{ij} \Bigg(\sum_{l=0}^{\infty} b_l n_{jk}^l\Bigg)
\label{eqn:power_series_corr}
\end{equation}

where $b_{l}$'s are arbitrary constants chosen in a way that $\sum_l b_l n_{jk}^l$ is converging and finite valued function. Using Eq.~\ref{eqn:std_dev_summand}, I get,

\begin{equation}
\tilde{g}_j = g_j - \sum_j \gamma_{ij} \Bigg(\sum_{l=0}^{\infty} b_l \frac{n_j^l}
{g_j^{l-1}} \Bigg)
\label{eqn:power_series_corr_modified}
\end{equation}

which removes all $k$-summands in Eq.~\ref{eqn:power_series_corr} and introduce indistinguishability in the model. Such elimination of $k$-summands is not possible for a similar power series expansion $n_{jk}$ with $l < 0$, hence those values are not allowed. Using the similar methods as described in the previous section, one can derive the optimal population distribution for self-correlating systems ($\gamma_{ij} \equiv \gamma\delta_{ij}$) as,

\begin{equation}
n_i = g_ie^{-(\alpha+\beta\epsilon_i)}\Bigg\{1-\gamma\Bigg(\sum_{l=0}^{\infty} (l+1)b_l\frac{n_i^l}{g_i^l}\Bigg)\Bigg\}
\label{eqn:power_series_corr_distribution}
\end{equation}

The maximum occupancy numbers at each degenerate state will have an upper bound for $\gamma > 0$ as,

\begin{equation}
0 \leq \sum_{l=0}^{\infty} (l+1)b_ln_c^l \leq \frac{1}{\gamma^+}
\label{eqn:max_occupancy_power_series}
\end{equation}

Using different values for the coefficients of $b_l$, one can derive different statistical models from Eq~\ref{eqn:power_series_corr_distribution}. Clearly, choice of $b_l$ has to be such that $b_l \to 0$ as $l \to \infty$ for $\gamma > 0$. 

An interesting example is $b_l = 1/(l+1)!$, which leads to,

\begin{equation}
n_i = g_ie^{-(\alpha+\beta\epsilon_i)}\Bigg\{1-\gamma e^{(n_i/g_i)}\Bigg\}
\label{eqn:exponential_distribution}
\end{equation}

As for dilute systems $\exp(n_i/g_i) \sim (1 + n_i/g_i)$, Eq.~\ref{eqn:exponential_distribution} roughly converges to Eq.~\ref{eqn:CFES} with $m=1$. In this particular case, one can show that, $\gamma^+(1+n_i/g_i) < 1$, which leads to, $\gamma^+ < 1$. 

%%%%%%%%%%%%%%%%%%%%%%%%%%%%%%%%%%%%%%%%%%%%%%%%%%%%%%%%%%%%%%%%%%%%%%%%%%%%%%%%
\section{Conclusion and Outlook}
\label{sect:conclusion}

The inter-dependency of statistical correlation and indistinguishability is an interesting aspect of this letter. Introducing a non-linear modification to the Haldane's theory of statistical correlation, I have shown that the special constraint used in the purely classical derivation of the CFES statistics is indeed linked to the indistinguishability of the particles. For a self-correlating system, by inserting the modified Haldane's theory into classical MB statistics directly produces CFES-equivalent intermediate statistics. Thus, I have shown that Haldane's statistical correlation theory lies at the core of CFES statistics. 

I have shown that one can derive many intermediate statistical models using appropriate choices for the $b_l$ coefficients, which will obey Eq.~\ref{eqn:max_occupancy_power_series} and produce physically realistic population distribution. An interesting question can be asked: Is self-correlation necessary to show quantum effects? Currently, I am exploring a binary inter-correlating (not self-correlating) system, which can answer this question and produce interesting results. The application of CFES statistics in enhanced sampling methods in molecular dynamics simulations of classical systems is also being investigated.

\acknowledgements
I am thankful to the National Institute of Technology, Tiruchirappalli, for providing the necessary funding for this research. 

%%%%%%%%%%%%%%%%%%%%%%%%%%%%%%%%%%%%%%%%%%%%%%%%%%%%%%%%%%%%%%%%%%%%%%%%%%%%%%%%
\bibliographystyle{eplbib}
\bibliography{CFES}
\end{document}